\documentclass[preprint,12pt]{elsarticle}

\usepackage{graphics}
\usepackage{graphicx}

\usepackage{epsfig}

\usepackage{amssymb}
\usepackage{amsthm}

\usepackage{lineno}
\usepackage{color}

\begin{document}

\begin{frontmatter}

 

\title{
von K\'arm\'an--Howarth and Corrsin equations closures through Liouville theorem
}
 

\author{Nicola de Divitiis}

\address{"La Sapienza" University, Dipartimento di Ingegneria Meccanica e 
Aerospaziale, Via Eudossiana, 18, 00184 Rome, Italy, \\
phone: +39--0644585268, \ \ fax: +39--0644585750, \\ 
e-mail: n.dedivitiis@gmail.com, \ \  nicola.dedivitiis@uniroma1.it}

\begin{abstract} 
In this communication, the closure formulas of von K\'arm\'an--Howarth and Corrsin equations 
are obtained through the Liouville theorem and the hypothesis of homogeneous isotropic incompressible turbulence.
Such closures, based on the concept that, in fully developed turbulence, contiguous fluid particles trajectories continuously diverge, are of non--diffusive nature, and express a correlations spatial propagation phenomenon between the several scales which occurs with a propagation speed depending on length scale and velocity standard deviation.
These closure formulas coincide with those just obtained in previous works through 
the finite scale Lyapunov analysis of the fluid act of motion.
Here, unlike the other articles, the present study does not use the Lyapunov theory, and
provides the closures showing first an exact relationship between the pair spatial correlations calculated with the velocity distribution function and those obtained using the material separation line distribution function. 
As this analysis does not adopt the Lyapunov theory, this does not need the definition and/or the existence of the Lyapunov exponents. Accordingly, the present proof of the closures results to be more general and rigorous than that presented in the other works, corroborating the previous results.
{\color{black}
Finally, the conditions of existence of invariants in isotropic turbulence are studied 
by means of the proposed closures. In the presence of such invariants and self--similarity, the sole evolution of velocity and temperature standard deviations and of the correlation scales is shown to be adequate to fairly describe the isotropic turbulence.
}
\end{abstract}

\begin{keyword}
Energy cascade, Bifurcations, Liouville theorem.
\end{keyword}

\end{frontmatter}

\newcommand{\no}{\noindent}
\newcommand{\be}{\begin{equation}}
\newcommand{\ee}{\end{equation}}
\newcommand{\bea}{\begin{eqnarray}}
\newcommand{\eea}{\end{eqnarray}}
\newcommand{\bc}{\begin{center}}
\newcommand{\ec}{\end{center}}

\newcommand{\calr}{{\cal R}}
\newcommand{\calv}{{\cal V}}

\newcommand{\bff}{\mbox{\boldmath $f$}}
\newcommand{\bfg}{\mbox{\boldmath $g$}}
\newcommand{\bfh}{\mbox{\boldmath $h$}}
\newcommand{\bfi}{\mbox{\boldmath $i$}}
\newcommand{\bfm}{\mbox{\boldmath $m$}}
\newcommand{\bfp}{\mbox{\boldmath $p$}}
\newcommand{\bfr}{\mbox{\boldmath $r$}}
\newcommand{\bfu}{\mbox{\boldmath $u$}}
\newcommand{\bfv}{\mbox{\boldmath $v$}}
\newcommand{\bfx}{\mbox{\boldmath $x$}}
\newcommand{\bfy}{\mbox{\boldmath $y$}}
\newcommand{\bfw}{\mbox{\boldmath $w$}}
\newcommand{\bfk}{\mbox{\boldmath $\kappa$}}

\newcommand{\bfA}{\mbox{\boldmath $A$}}
\newcommand{\bfD}{\mbox{\boldmath $D$}}
\newcommand{\bfI}{\mbox{\boldmath $I$}}
\newcommand{\bfL}{\mbox{\boldmath $L$}}
\newcommand{\bfM}
{\mbox{\boldmath $M$}}
\newcommand{\bfS}{\mbox{\boldmath $S$}}
\newcommand{\bfT}{\mbox{\boldmath $T$}}
\newcommand{\bfU}{\mbox{\boldmath $U$}}
\newcommand{\bfX}{\mbox{\boldmath $X$}}
\newcommand{\bfY}{\mbox{\boldmath $Y$}}
\newcommand{\bfK}{\mbox{\boldmatthe average of $u_\xi u_\xi^*/u^2$h $K$}}

\newcommand{\bfeta}{\mbox{\boldmath $\eta$}}
\newcommand{\bfrho}{\mbox{\boldmath $\rho$}}
\newcommand{\bfchi}{\mbox{\boldmath $\chi$}}
\newcommand{\bfphi}{\mbox{\boldmath $\phi$}}
\newcommand{\bfPhi}{\mbox{\boldmath $\Phi$}}
\newcommand{\bflambda}{\mbox{\boldmath $\lambda$}}
\newcommand{\bfxi}{\mbox{\boldmath $\xi$}}
\newcommand{\bfLambda}{\mbox{\boldmath $\Lambda$}}
\newcommand{\bfPsi}{\mbox{\boldmath $\Psi$}}
\newcommand{\bfomega}{\mbox{\boldmath $\omega$}}
\newcommand{\bfOmega}{\mbox{\boldmath $\Omega$}}
\newcommand{\bfeps}{\mbox{\boldmath $\varepsilon$}}
\newcommand{\bfepsn}{\mbox{\boldmath $\epsilon$}}
\newcommand{\bfzeta}{\mbox{\boldmath $\zeta$}}
\newcommand{\bfkappa}{\mbox{\boldmath $\kappa$}}
\newcommand{\bfsigma}{\mbox{\boldmath $\sigma$}}
\newcommand{\itPsi}{\mbox{\it $\Psi$}}
\newcommand{\itPhi}{\mbox{\it $\Phi$}}

\newcommand{\bint}{\mbox{ \int{a}{b}} }
\newcommand{\ds}{\displaystyle}
\newcommand{\Sum}{\Large \sum}



\bigskip

\section{Introduction \label{intro}}

The von K\'arm\'an--Howarth and Corrsin equations are evolution equations of longitudinal velocity and temperature correlations in homogeneous isotropic turbulence, respectively.
Both the equations, being unclosed, need the adoption of proper closures 
\cite{Karman38, Batchelor53, Corrsin_1, Corrsin_2}. In detail, such equations include $K$ and $G$, terms due, respectively, to inertia forces and to temperature convective effect. These terms, directly related to the longitudinal triple velocity correlation and to the triple velocity--temperature correlation, require an adequate modeling which must take into account that inertia forces and convective effect do not modify average kinetic and thermal energies. In addition $K$ satisfies the detailed conservation of energy \cite{Batchelor53} following which the exchange of energy between wave--numbers is only linked to the amplitudes of such wave--numbers and of their difference \cite{Eyink06}.
Although numerous articles were written which concern the closures of the autocorrelation equations in the Fourier domain \cite{Obukhov41, Heisenberg48, Kovasznay48, Ellison61, Pao65, Leith67, Clark98, Nazarenko04, Clark09}, few articles address the closures of $K$ and $G$ in the physical space \cite{Hasselmann58, Millionshtchikov69, Oberlack93, Baev}. These latter, being based on the eddy--viscosity concept, describe diffusive closure models.
In such framework, Hasselmann \cite{Hasselmann58} proposed, in 1958, a closure model which expresses $K$ through a complex expression, and Millionshtchikov developed a closure model which exhibits an empirical constant \cite{Millionshtchikov69}.
Although both these models propose two closures that in particular conditions adequately describe the  energy cascade, in general, these do not satisfy some physical conditions.
More recently, Oberlack and Peters \cite{Oberlack93} suggested a closure that exhibits a free  parameter. The authors show that, for a proper choice of such parameter, the closure reproduces the energy cascade providing results in agreement with the experiments \cite{Oberlack93}.
Thereafter, Khabirov and Unal \cite{Khabirov1, Khabirov2} studied the non--closed von K\'arm\'an--Howarth equation by group theoretical methods and suggested solutions to the closure problem of isotropic turbulence, especially for the decay of the turbulence.

For what concerns the Corrsin equation, Baev and Chernykh \cite{Baev} (and references therein) analyzed velocity and temperature correlations by means of a closure model based on the gradient hypothesis which relates pair longitudinal second and third order correlations, by means of empirical coefficients.
Although other works regarding the von K\'arm\'an--Howarth and Corrsin equations were written 
\cite{Mellor84, Onufriev94, Grebenev05, Grebenev09, Antonia2013}, \cite{George1, George2, Antonia}, to the author's knowledge a physical--mathematical analysis based on basic statistical principles which lead to the analytical closures of these equations has not received due attention.
Therefore, the purpose of this article is to propose closures of the von K\'arm\'an--Howarth and Corrsin equations by means of a theory which does not adopt the eddy--viscosity paradigm.

Unlike the other works, the proposed analysis provides nondiffusive closures in the physical space
which are based on the Liouville theorem, and on the concept that, due to bifurcations in developed turbulence, contiguous fluid particles trajectories continuously  diverge. Specifically, the proposed closures are obtained through the rate of material separation line distribution function in isotropic turbulence whose analytical form is achieved by means of the Liouville theorem. Such closures 
correspond to a spatial propagation mechanism of the correlations between the scales which happens with a propagation speed depending on length scale and velocity standard deviation. 
These formulas coincide with those just obtained in Refs. \cite{deDivitiis_1, deDivitiis_4} and \cite{deDivitiis_5} where the author shows that these closures adequately describe the energy cascade phenomenon, reproducing negative skewness of velocity difference in very good agreement with the litarature data \cite{Chen92, Orszag72, Panda89, Anderson99, Carati95, Kang2003}, providing the Kolmogorov law and temperature spectra in line with the theoretical argumentation of Kolmogorov, Obukhov--Corrsin and Batchelor \cite{Batchelor_2, Batchelor_3, Obukhov}, with experimental results \cite{Gibson, Mydlarski}, and with numerical data \cite{Rogallo, Donzis}.

Unlike the previous articles  \cite{deDivitiis_1, deDivitiis_4, deDivitiis_5, deDivitiis_8}, which derive such closures 
through finite scale Lyapunov theory and Liouville theorem, the present analysis does not adopt the Lyapunov theory. The present formulation only uses Liouville theorem and distribution function rate of the fluid placement.

Specifically, the present proof of the closures is based on the following elements: 

a) As the consequence of bifurcations properties and fully developed chaos hypothesis, the material separation lines distribution function is assumed to be statistically independent of the velocity field distribution function. 

b) The material line distribution function is expressed by means of the hypotheses of fluid incompressibility and statistical isotropy.

c) Due to statistical isotropy, an exact relation is first obtained between spatial correlations calculated using the distribution function of material line and those obtained by means of the velocity field distribution function. This relation, necessary for obtaining the proposed closures, represents a novelty of the present work with respect to the previous ones where such relationship was assumed to be valid in an intuitive way.

d) Next, thanks to their properties of maintaining unaltered average kinetic and thermal energies,
 $K$ and $G$ are recognized to be only due to the rate of separation lines distribution function,
where the latter is formally expressed by means of the Liouville theorem.

Hence, the present analysis does not require the Lyapunov theory and results to be more general and  rigorous than the formulations presented in \cite{deDivitiis_1, deDivitiis_4, deDivitiis_5, deDivitiis_8}, showing that the Lyapunov theory is a sufficient analytical tool for achieving the same formulas,  corroborating the previous results. 

{\color{black}
Thereafter, the conditions of existence of invariants are analyzed using the proposed closures, with particular reference to the integrals of Loitsianskii and of Saffman--Birkhoff \cite{Loitsianskii39, Saffman67} and to self--similarity produced by the same closure formulas. We show that,  in the presence of self--similarity and invariants, the sole evolution of velocity and temperature standard deviations and of the correlation scales provides a fair description of isotropic turbulence.
}

\begin{figure}
	\centering
	\includegraphics[width=80mm,height=110mm]{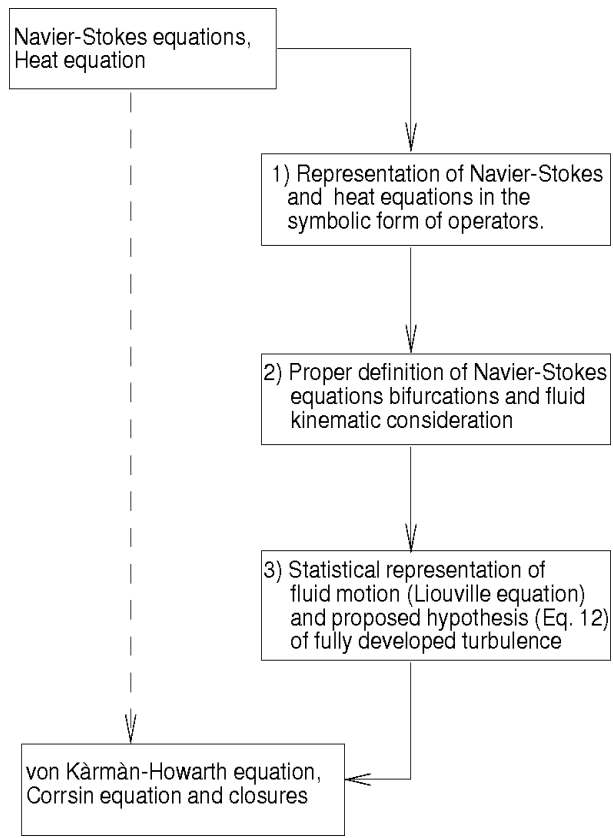}
\caption{Scheme of the proposed analysis.}
\label{figura_1}
\end{figure}

{\color{black} We conclude this section by remarking the way in which the proposed closures are obtained. These formulas are not achieved by passing directly from the Navier--Stokes and heat equations to the correlations equations (dashed line, see Fig. 1) for instance by means of phenomenological hypotheses. Such closures are here determined according to specific analytical formulation and hypotheses reported in the boxes 1), 2), 3) of the flowchart (solid line) and presented in this work.}

\bigskip

\section{Background \label{Background}}

This background has the purpose to summarize the link between Navier--Stokes equations bifurcations, fluctuations of velocity and temperature fields, and fluid particles trajectories divergence, which is useful for the present analysis.
To this end, consider now the Navier--Stokes equations and heat equation 
\bea
\begin{array}{l@{\hspace{-0.cm}}l}
\ds \nabla_{\bf x} \cdot {\bf u} =0, \\\\
\ds \frac{\partial {\bf u}}{\partial t} =
-  \nabla_{\bf x} {\bf u} \ {\bf u} - \frac{\nabla_{\bf x}p}{\rho}  + \nu \nabla_{\bf x}^2 {\bf u}
\label{NS_eq}
\end{array}
\eea
\bea
\begin{array}{l@{\hspace{-0.cm}}l}
\ds \frac{\partial \vartheta }{\partial t} =
-{\bf u} \cdot \nabla_{\bf x} \vartheta + \chi \nabla_{\bf x}^2 \vartheta
\label{T_eq}
\end{array}
\eea
 $\bf u$=${\bf u}(t, {\bf x})$, $p$=$p(t, {\bf x})$ and $\vartheta$=$\vartheta(t, {\bf x})$
are velocity, pressure and temperature fields, $\nu$ and $\chi$=$k \rho/C_p$ are fluid kinematic viscosity and thermal diffusivity, being $\rho=$const, $k$ and $C_p$ density, fluid thermal conductivity and specific heat at constant pressure, respectively. 
In this study $\nu$ and $\chi$ are supposed to be independent from the temperature, thus Eqs. (\ref{NS_eq}) is autonomous with respect to Eq. (\ref{T_eq}), whereas Eq. (\ref{T_eq}) will depend on Eqs. (\ref{NS_eq}).

The bifurcations of the partial differential system (\ref{NS_eq})--(\ref{T_eq}) are propely defined in line with Refs. \cite{deDivitiis_5, deDivitiis_8}, reducing first Eqs. (\ref{NS_eq})--(\ref{T_eq}) to the symbolic form of operators. In detail,  the pressure is eliminated in the momentum Navier--Stokes equations through the continuity equation, therefore Eqs. (\ref{NS_eq}) and (\ref{T_eq}) formally read as follows
\bea
\begin{array}{l@{\hspace{-0.cm}}l}
 \dot{\bf u} =  {\bf N}({\bf u} ; \nu),
\end{array}
\label{NS_op}
\eea
\bea
\begin{array}{l@{\hspace{-0.cm}}l}
\dot{\vartheta} =  {\bf M}({\bf u}, \vartheta ; \chi)
\end{array}
\label{T_op}
\eea
wherein ${\bf N}$ is a quadratic operator which includes, among the other terms,
the integral nonlinear operator which provides $\nabla_{\bf x} p$ as functional of the velocity field, being 
\bea
\ds  p(t, {\bf x}) =  \frac{\rho}{4 \pi}
\int \frac{\partial^2 u_i' u_j'}{\partial x_i' \partial x_j'} \ \frac{d V({\bf x}')}{\vert {\bf x}' - {\bf x} \vert} 
\label{pressure}
\eea
The pressure, being a functional of {\bf u}, produces nonlocal effects \cite{Tsinober2009}, and the Navier--Stokes equations, reduced to be an integro--differential equation, is formally given by Eq. (\ref{NS_op}) in the symbolic form of operators. 
For what concerns Eq. (\ref{T_op}), it is the evolution equation of $\vartheta$, where $\bf M$ is a linear operator of $\vartheta$. 

At this stage of the present study, Eqs. (\ref{NS_op}) and (\ref{T_op}) can be dealt with according to the analysis of Ref. \cite{Ruelle71} where the authors supposed that the infinite dimensional space of velocity field $\left\lbrace \bf u \right\rbrace$ can be replaced by a finite-–dimensional manifold. Hence, Eqs. (\ref{NS_op}) and (\ref{T_op}) can be reduced to be one equation of the kind studied by Ruelle and Takens in Ref. \cite{Ruelle71}, and the classical bifurcation theory of ordinary differential equations \cite{Ruelle71, Eckmann81,  Guckenheimer90} can be formally applied to Eq. (\ref{NS_op}) and (\ref{T_op}). This can be considered to be valid in the limits of the formulation proposed in Ref. \cite{Ruelle71}.
As $\bf M$ is linear with respect to $\vartheta$, transition and turbulence are caused by the bifurcations of Eq. (\ref{NS_op}), where $\nu^{-1}$ plays the role of the control parameter. 
Such bifurcations occur in the points of $\left\lbrace \bf u \right\rbrace$ where the Jacobian ${\nabla_{\bf u} {\bf N}}$ exhibits at least an eigenvalue with zero real part (NS--bifurcations), and this occurs when
\bea
\ds \det (\nabla_{\bf u} {\bf N}) =0.
\eea
Such bifurcations are responsible for multiple velocity fields $\hat {\bf u}$ which correspond to
the same field $\dot{\bf u}$.
In fact, during the fluid motion, multiple solutions $\hat {\bf u}$ and $\hat{\vartheta}$ can be determined, at each instant, through inversion of Eq. (\ref{NS_op})
\bea
\begin{array}{l@{\hspace{-0.cm}}l}
\dot{\bf u} = {\bf N}({\bf u}; \nu)  \\\\
\hat {\bf u}  = {\bf N}^{-1}(\dot{\bf u}; \nu), \\\\
\hat{\vartheta}=  {\bf M}^{-1}(\dot{\vartheta}, \hat {\bf u}; \chi)
\end{array}
\label{invr0}
\eea
One single Navier--Stokes bifurcation, which produces doubling of the
values of $\bf u$, also causes doubling of all the characteristics associated with velocity and
temperature fields, with particular reference to their characteristic scales \cite{deDivitiis_8}.
Therefore, in fully developed turbulence, the number of the Navier--Stokes equations bifurcations diverges, and velocity and temperature fields and their length scales are continuously distributed \cite{deDivitiis_8}.

On the other hand, the fluid volume changes its placement and deforms its shape
according to
\bea
\begin{array}{l@{\hspace{-0.cm}}l}
\ds \dot{{\bfx} } = {\bf u} (t, {\bfx}),   \\\\
\ds \dot{{\bfxi} } = {\bf u} (t, {\bfx}+{\bfxi}) - {\bf u} (t, {\bfx}),
\end{array}
\label{kin finite}
\eea
Equations (\ref{kin finite}) provide fluid displacement evolution and relative kinematics,
where ${\bfx}(t)$ and ${\bfy}(t)={\bfx}(t)+{\bfxi}(t)$ are two arbitrary fluid particles trajectories, and  $\bfxi$ is the corresponding separation vector.
One point of the physical space is of bifurcation for the velocity field (kinematic bifurcation) if ${\nabla_{\bf x} {\bf u}}(t, {\bf x})$ has at least an eigenvalue with zero real part, and this happens when its determinant vanishes, i.e.
\bea
\ds \det \left( \nabla_{\bf x} {\bf u} (t, {\bf x}) \right) = 0.
\eea
Now, the Navier--Stokes bifurcations have significant implications for what concerns the relative kinematics of velocity field. Specifically, Ref. \cite{deDivitiis_5, deDivitiis_8} show that, the continuous doubling of velocity field values and of the corresponding scales cause non smooth spatial variations of ${\bf u}(t, {\bf x})$, $t>$0, which in turn deternine very frequent kinematic bifurcations, and that, in fully developed turbulence, the fluctuations of $\bfxi$ are much more rapid and statistically independent with respect to the time variations of velocity field.
Due to the fluid incompressibility, two fluid particles will describe chaotic trajectories, 
${\bfx}(t)$ and ${\bf y}(t)={\bfx}(t)+{\bfxi}(t)$, which diverge with each other with a local rate of divergence quantified by the local velocity longitudinal component ${U}$
\bea
\ds {U} = \dot{\bfxi} \cdot \frac{\bfxi}{\xi}
\label{U}
\eea
Because of its definition (\ref{U}) and according to the analysis of Ref. \cite{deDivitiis_5, deDivitiis_8}, $U$ is a fluctuating quantity much faster than $\bf u$ and $\vartheta$, whose distribution function directly arises from the statistics of $\bfxi$.

{ Accordingly, in fully developed turbulence, the time--scales of $\bfxi$ are expected to be completely separated from those associated with $\bf u$ and $\vartheta$ in the sense that 
$\bfxi$ and ($\bf u$, $\vartheta$) exhibit chaotic behavior and their power spectra 
are supposed to be located in frequency intervals which are completely separated \cite{deDivitiis_3}. 
Thus, $\bfxi$ and ($\bf u$, $\vartheta$) are considered to be statistically uncorrelated.
This means that the effect of the trajectories divergence is much more rapid and statistically uncorrelated with respect to the variations of the velocity field.
This property is supported by the arguments presented in Refs. \cite{Ottino89, Ottino90} (and references therein), where the author observes the that:
a) The fields ${\bf u} (t, {\bfx})$, (and therefore also ${\bf u} (t, {\bfx+ \bfxi})- {\bf u} (t,  {\bfx})$) produce chaotic trajectories also for relatively simple mathematical structure of the right--hand sides ${\bf u} (t, {\bfx})$ (also for steady fields!). 
b) The flows given by ${\bf u} (t, {\bfx})$ (and therefore also ${\bf u} (t, {\bfx+ \bfxi})- {\bf u} (t, {\bfx})$) stretch and fold continuously and rapidly causing an effective mixing of the particles trajectories.
}

\bigskip

\section{Distribution functions of ${\bf u}$, $\vartheta$, $\bfx$, $\bfxi$ and $U$}

According to this analysis, ${\bf u}$, $\vartheta$, $\bfx$ and $\bfxi$ are 
fluid state variables, thus, {\color{black} in the framework of Liouville theorem for nonlinear equations \cite{Nicolis95, Wei19}}, the distribution function of these quantities, say 
$P$, follows the Liouville equation associated with Eqs. (\ref{NS_op})--(\ref{T_op}) and (\ref{kin finite}) 
\bea
\begin{array}{l@{\hspace{-0.cm}}l}
\ds \frac{\partial P}{\partial t} +
\frac{\delta}{\delta {\bf u}} \cdot \left( P \dot{\bf u} \right) +
\frac{\delta}{\delta \vartheta} \cdot \left( P \dot{\vartheta} \right) + 
\frac{\partial}{\partial {\bfx}} \cdot \left( P \dot{\bfx} \right) +
\frac{\partial}{\partial {\bfxi}} \cdot \left( P \dot{\bfxi} \right) 
=0
\end{array}
\label{Liouville}
\eea
where, following the notation of Eqs. (\ref{NS_op})--(\ref{T_op}), 
$\delta/ \delta {\bf u}$ and  $\delta/ \delta {\vartheta}$ are functional partial derivatives
with respect to $\bf u$ and $\vartheta$, respectively and 
$\left( \partial/ \partial \circ \right) \cdot$ stands for the divergence with respect to $\circ$.
In line with the previous sections and with Refs. \cite{deDivitiis_5, deDivitiis_6, deDivitiis_8}, $P$ can be factorized as follows
\bea 
\begin{array}{l@{\hspace{-0.cm}}l}
\ds P(t, {\bf u}, \vartheta, {\bfx}, {\bfxi}) = F(t, {\bf u}, \vartheta) P_\xi(t, {\bfx}, {\bfxi})
\end{array}
\label{indep stat}
\eea
being $F$ and $P_\xi$ the distribution functions of ($\bf u$, $\vartheta$), and of ($\bfx$, $\bfxi$), respectively.
Observe that Eq. (\ref{indep stat}) represents the crucial point of this study, being the hypothesis of fully developed turbulence of this analysis.
The instantaneous rates of $F$ and $P_\xi$ are formally obtained from Eq. (\ref{Liouville}), taking into account the aforementioned statistical independence (\ref{indep stat}).
On the other hand, in homogeneous isotropic turbulence, $P_\xi$ does not depend on $\bfx$, and results to be a function of $\vert \bfxi \vert -r$,  where $r$ plays the role of finite scale. Moreover, as $\vert \bfxi \vert$ is considered to be given at each instant, $P_\xi$ is a distribution of $\bfxi$ strongly peaked in $\vert \bfxi \vert =r$. 
This implies that, at each instant, in homogenous isotropic turbulence in infinite fluid region, the 
distribution function of $\bfxi$ can be represented by the following form
\bea
\begin{array}{l@{\hspace{-0.cm}}l}
\ds P_\xi= 
 \frac{1}{4 \pi r^2} \ \delta(\xi -r), \\\\
\ds \xi = \vert \bfxi \vert, \\\\
\ds \forall t \geq 0
\end{array}
\label{Pxi_1} 
\eea
and its variation rate effects can be formally calculated by means of Eq. (\ref{Liouville}).

Also $U$ is statistically independent of the velocity field and is continuously distributed in its range of variation, being
\bea
\begin{array}{l@{\hspace{-0.cm}}l}
\ds U \in \left(U_m, U_S \right), \\\\
\ds U_m = \inf\left\lbrace U \right\rbrace, \ \ \ U_S =\sup\left\lbrace  U \right\rbrace 
\end{array}
\eea
In order to determine the interval $\left(U_m, U_S \right)$ in incompressible isotropic turbulence, consider first
the volume built on the separation vectors $\bfxi_1$, $\bfxi_2$ and $\bfxi_3$ initially orthogonal with each other such that $\vert \bfxi_1 \vert$ =$\vert \bfxi_2 \vert$=$\vert \bfxi_3 \vert$=$\xi$, which satisfy Eq. (\ref{kin finite}). This volume can be expressed by 
\bea
\ds {\cal V} = \left( {\bfxi}_1 \times {\bfxi}_2 \right) \cdot {\bfxi}_3
\eea
thus, its rate reads as follows
\bea
\begin{array}{l@{\hspace{-0.cm}}l}
\ds \frac{d{\cal V}}{dt} = 
\left( \dot{\bfxi}_1 \times {\bfxi}_2 \right) \cdot {\bfxi}_3+ 
\left( {\bfxi}_1 \times \dot{\bfxi}_2\right)  \cdot {\bfxi}_3 + 
\ds \left( {\bfxi}_1 \times {\bfxi}_2 \right) \cdot \dot{\bfxi}_3
\end{array}
\eea
If these vectors are infinitesimal in the sense that $\xi \rightarrow 0$, due to continuity
 ${d{\cal V}}/{dt}\equiv$0. On the contrary, when $\xi$ is finite, the rate of $\cal V$ can be 
different from zero. Nevertheless, due to homogeneous and isotropic turbulence, 
the average of ${d{\cal V}}/{dt}$ is expected to be equal to zero, i.e.
\bea
\begin{array}{l@{\hspace{-0.cm}}l}
\ds \left\langle \frac{d{\cal V}}{dt} \right\rangle_\xi = 
\left\langle \left( \dot{\bfxi}_1 \times {\bfxi}_2 \right) \cdot {\bfxi}_3+ 
\left( {\bfxi}_1 \times \dot{\bfxi}_2\right)  \cdot {\bfxi}_3 + 
\left( {\bfxi}_1 \times {\bfxi}_2 \right) \cdot \dot{\bfxi}_3\right\rangle_\xi =0
\end{array}
\label{ave dv/dt}
\eea
where $\left\langle \circ \right\rangle_\xi$ denotes the average of $\circ$ calculated, through
$P_\xi$. Adopting the longitudinal components of velocity differences $\dot{\bfxi}_k$, defined as 
\bea
\begin{array}{l@{\hspace{-0.cm}}l}
\ds  \ U_k = \dot{\bfxi}_k \cdot \frac{\bfxi_k}{\xi_k}, \ \ k = 1, 2, 3
\end{array}
\label{U_k}
\eea
Eq. (\ref{ave dv/dt}) can be written in terms of $U_k$ in the following manner
\bea
\begin{array}{l@{\hspace{-0.cm}}l}
\ds \left\langle U_1 +U_2 + U_3 \right\rangle_\xi \xi^2  =0, 
\end{array}
\label{u_1+u_2+u_3}
\eea
where, because of statistical hypothesis of isotropy, each addend of Eq. (\ref{u_1+u_2+u_3})
can be expressed in such a way that 
\bea
\begin{array}{l@{\hspace{-0.cm}}l}
\ds U_k  =  A_k  \cos\left( \varepsilon+ \frac{2}{3} \left( k-1\right)  \pi\right), \ \ k = 1, 2, 3, \\\\
\ds   \left\langle A_1 \right\rangle_\xi = \left\langle A_2 \right\rangle_\xi= \left\langle A_3 \right\rangle_\xi, \\\\
\ds   \sup\left\lbrace  A_1 \right\rbrace = \sup\left\lbrace A_2 \right\rbrace = \sup\left\lbrace A_3 \right\rbrace \equiv A_S, 
\end{array}
\label{u_k}
\eea
in which $\varepsilon$ has to be determined, whereas $A_k$ play the role of fluctuating variables.
Now, according to the hypothesis of fully developed chaos, the number of kinematic bifurcations diverges, thus it is expected that $U_S \equiv \sup \left\lbrace U \right\rbrace \ge $0 will assume its maximum value compatible with Eq.(\ref{u_k}), whereas $U_m \equiv \inf \left\lbrace U \right\rbrace \le $0 is   consequently determined. 
Therefore, $U_S$ is calculated as
\bea
\begin{array}{l@{\hspace{-0.cm}}l}
\ds U_S = \sup_{A_k =A_S, \ \varepsilon } \left\lbrace U_1, U_2, U_3 \right\rbrace \\\\
\ds \equiv \sup_{\varepsilon} \left\lbrace \left\lbrace U_1 \right\rbrace_{A_1=A_S}, \left\lbrace U_2 \right\rbrace_{A_2=A_S}, \left\lbrace U_3 \right\rbrace_{A_3=A_S}  \right\rbrace \ge 0, 
\end{array}
\eea
and the corresponding value of $\varepsilon$, say $\varepsilon_S$, is then formally written as
\bea
\begin{array}{l@{\hspace{-0.cm}}l}
\ds \varepsilon_S = \mbox{argmax} \left\lbrace \sup \left\lbrace \left\lbrace U_1\right\rbrace_{A_1=A_S}, \left\lbrace U_2 \right\rbrace_{A_2=A_S}, \left\lbrace U_3 \right\rbrace_{A_3=A_S}  \right\rbrace \right\rbrace 
\end{array}
\eea
This implies that $\varepsilon_S=2 (k-1) \pi$, $k$=1, 2,..., and that $U_m$ is
\bea
\begin{array}{l@{\hspace{-0.cm}}l}
\ds U_m = \inf_{\varepsilon=\varepsilon_S} \left\lbrace 
\left\lbrace U_1\right\rbrace_{A_1=A_S}, \left\lbrace U_2 \right\rbrace_{A_2=A_S}, \left\lbrace U_3 \right\rbrace_{A_3=A_S}
\right\rbrace = -\frac{U_S}{2} \le 0, 
\end{array}
\eea
That is
\bea
\ds U \in \left( -\frac{U_S}{2}, U_S\right) 
\label{U range}
\eea

Hence, the distribution function of $U$, say $P_U$, can be formally expressed by means of $P_\xi$ through the Frobenius--Perron equation \cite{Nicolis95}
\bea
\ds P_U \left( U \right)
=  \int_{\Xi} P_\xi \ \delta\left( U- \frac{\dot{\bfxi}\cdot {\bfxi}}{\xi}\right)  \  d{\Xi} 
\label{FrobeniusPerron}
\eea
With reference to Eq. (\ref{FrobeniusPerron}), when ${\bfxi}$ sweeps the sphere $\vert {\bfxi} \vert =r$, $U$ describes the corresponding interval given by Eq. (\ref{U range}).
In isotropic turbulence there are no privileged directions, thus the longitudinal component of the velocity difference $\dot{\bfxi} \cdot{\bfxi}/\xi$ results to be uniformely distributed in its variation range as ${\bfxi}$ varies in such a way that  $\vert {\bfxi} \vert =r$. 
The distribution function of $U$ is then formally obtained substituting Eq. (\ref{Pxi_1}) in Eq. (\ref{FrobeniusPerron})  
\bea
\ds P_U = 
\left\lbrace 
\begin{array}{l@{\hspace{-0.cm}}l}
\ds \frac{2}{3}\frac{1}{U_S}, \ \ \mbox{if} \  U \in \left( -\frac{U_S}{2}, U_S\right)  \\\\
\ds 0 \ \ \mbox{elsewhere} 
\end{array}\right. 
\label{Pl}
\eea
Such distribution function provides $\langle U \rangle_\xi >0$ and, among the other properties, gives in particular the link between $\langle U \rangle_\xi$ and $\langle U^2 \rangle_\xi$, according to 
\bea
\begin{array}{l@{\hspace{-0.cm}}l}
\ds \left\langle U \right\rangle_\xi =
\frac{1}{2} \sqrt{\left\langle U^2 \right\rangle_\xi} > 0.
\end{array}
\label{mPl}
\eea 
We conclude this section by observing that, the distribution (\ref{Pl}) is the result of two elements: fully developed chaos and fluid incompressibility. The fully developed chaos produces, in any case, the  particles trajectories divergence, and, in the particular case of infinite fluid domain, also the flow statistical isotropy. The combined effect of trajectories divergence and fluid incompressibility determines the domain of $P_U$: the trajectories divergence, representing element of instability, is responsible for the variation intervals where $U>$0, whereas the incompressibility acts in opposite sense preserving the volume and determining regions where $U<$0. Next, the statistical isotropy, providing no privileged directions, causes an uniform distribution of $U \in (-U_S/2, U_S)$.
These results are in agreement with those just obtained in Ref. \cite{deDivitiis_6}, where, the author, adopting the Lyapunov theory, shows that the finite scale Lyapunov exponent associated with 
Eq. (\ref{kin finite}) results to be uniformely distributed in its variation range.

\bigskip

\section{Relation between spatial correlations in isotropic turbulence}

One of the consequences of the hypotheses of fully developed turbulence and statistical isotropy is the link between the statistical spatial correlations calculated through $P_\xi$ and those obtained by means of $F$.
Such property, in conjunction with the previous analysis, leads
to the analytical forms of the closures of the von K\'arm\'an-Howarth and Corrsin equations.

Among the various statistical spatial correlation functions, consider for our pusposes the pair correlation of the longitudinal components of velocity $\left\langle u_r u_r' \right\rangle$.
This can be calculated in terms of $F$ as follows
\bea
\begin{array}{l@{\hspace{-0.cm}}l}
\ds \left\langle u_r u_r' \right\rangle \equiv \int_{{\cal U}}  F \ u_r u_r' \ d {\cal U}
\end{array}
\label{corr u}
\eea
where $\left\langle \circ \right\rangle$ denotes the average of $\circ$ calculated, through $F$, 
 $\ds {\cal U} = \left\lbrace {\bf u}\right\rbrace \times \left\lbrace {\vartheta}\right\rbrace$,
$d{\cal U}$ stands for the corresponding elemental volume, and
\bea
\begin{array}{l@{\hspace{-0.cm}}l}
\ds u_r = {\bf u}(t, {\bf x})\cdot \frac{\bf r}{r}, \ u_r' = {\bf u}(t, {\bf x}+ {\bf r})\cdot \frac{\bf r}{r}, \ 
\end{array}
\eea
On the other hand, following the present analysis, in fully developed turbulence, the kinematic bifurcations diverge.
This result and the hypothesis of statistical isotropy allow to analytically express $\left\langle u_r u_r' \right\rangle$ as the average of $u_\xi u_\xi^*$ calculated as the surface integral over the spherical surface $S({\bfxi})$: $\xi =r$, i.e.
\bea
\begin{array}{l@{\hspace{-0.cm}}l}
\ds \left\langle u_r u_r' \right\rangle = \frac{1}{4 \pi r^2} \int_{S\left({\xi}\right)} u_\xi u_\xi^* \ d S({\bfxi}),
\end{array}
\label{urur'}
\eea 
being
\bea
\begin{array}{l@{\hspace{-0.cm}}l}
\ds u_\xi = {\bf u}(t, {\bf x})\cdot \frac{\bfxi}{\xi}, \ u_\xi^* = {\bf u}(t, {\bf x}+ {\bfxi})\cdot \frac{\bfxi}{\xi}, \ 
\end{array}
\eea 
Now, from Eq. (\ref{Pxi_1}), the surface integral of Eq. (\ref{urur'}) is the layer integral calculated through $P_\xi$ following the Minkowski content measure \cite{Federer69}. Hence, in isotropic fully developed turbulence, the correlation $\left\langle u_r u_r' \right\rangle$ can be also expressed in terms of $P_\xi$ as follows
\bea
\begin{array}{l@{\hspace{-0.cm}}l}
\ds \left\langle u_r u_r' \right\rangle = \int_{\Xi}  P_\xi \ u_\xi u_\xi^* \ d {\Xi} 
\equiv \left\langle u_\xi u_\xi^* \right\rangle_\xi, \ \forall r > 0
\end{array}
\label{corr u 2}
\eea
where $\Xi\equiv \left\lbrace {\bfxi}\right\rbrace$ and $d \Xi$ denotes the corresponding elemental
volume.
More in general, a link of the kind of Eq. (\ref{corr u 2}) will regard the several spatial correlations  built through velocity and/or temperature fields with both the distribution functions. 
This represents the novelty of the present work with respect the previous ones. While in the previous articles, Eq. (\ref{corr u 2}) is assumed to be hold intuitively, here this equation is obtained by means of the isotropy hypothesis and of the properties of $P_\xi$.

It is worth to remark that Eq. (\ref{corr u 2}) holds only when $r$ is strictly positive and when the kinematic bifurcations diverge. On the contrary, in case of non developed turbulence, such as during
the transition through intermediate stages of turbulence, or in more complicate situations with boundary conditions, for instance near the wall, Eq. (\ref{corr u 2}) cannot be applied.

In conclusion, although $F$ and $P_\xi$ are two different and independent distribution functions, the hypotheses of fully developed turbulence and statistical isotropy relate the various spatial correlations calculated through $F$ to those achieved by means of $P_\xi$. 

\bigskip

\section{Closure of von K\'arm\'an--Howarth and Corrsin equations}

The closures of von K\'arm\'an-Howarth and Corrsin equations are here obtained 
using the Liouville theorem , the statistical independence (\ref{indep stat}), 
and the properties of $P_\xi$ given by Eqs. (\ref{corr u 2}) and (\ref{property}) (see  Appendix).
These equations, properly obtained from the Navier--Stokes and heat equations written in two points of space, $\bf x$ and $\bf x'= x + r$, read as follows 
\bea
\begin{array}{l@{\hspace{-0.cm}}l}
\ds \frac{\partial f}{\partial t} = 
\ds  \frac{K}{u^2} +
\ds 2 \nu  \left(  \frac{\partial^2 f} {\partial r^2} +
\ds \frac{4}{r} \frac{\partial f}{\partial r}  \right) +\frac{10 \nu}{\lambda_T^2} f, \\\\
\ds \frac{\partial f_\theta}{\partial t} = 
\ds  \frac{G}{\theta^2} +
\ds 2 \chi  \left(  \frac{\partial^2 f_\theta} {\partial r^2} +
\ds \frac{2}{r} \frac{\partial f_\theta}{\partial r}  \right) +\frac{12 \chi}{\lambda_\theta^2} f_\theta,
\end{array}
\label{vk-h}
\eea
whose boundary conditions are
\bea
\begin{array}{l@{\hspace{+0.2cm}}l}
\ds f(0) = 1,  \ \ \ \ \ds \lim_{r \rightarrow \infty} f (r) = 0, \\\\
\ds f_\theta(0) = 1,  \ \ \ \ \ds \lim_{r \rightarrow \infty} f_\theta (r) = 0, 
\end{array}
\label{bc0}
\eea
where $f=\langle u_r u_r' \rangle/u^2$ and $f_\theta = \langle \vartheta  \vartheta' \rangle/\theta^2$ are the pair correlations of longitudinal velocity components and of temperature, 
$u \equiv \sqrt{\langle u_r^2 \rangle}$,  $\theta \equiv \sqrt{\langle \vartheta^2 \rangle}$, being $\lambda_T \equiv \sqrt{-1/f''(0)}$ and $\lambda_\theta \equiv \sqrt{-2/f_\theta''(0)}$ Taylor and Corrsin microscales, respectively.
$K$ and $G$, arising from inertia forces and convective terms, give the energy cascade, and are in terms of velocity and temperature fields according to  \cite{Karman38, Corrsin_1, Corrsin_2} 
\bea
\begin{array}{l@{\hspace{-0.cm}}l}
\ds \left( 3 + r \frac{\partial}{\partial r} \right) K = 
\frac{\partial}{\partial r_k} \left\langle  u_i u_i'\left( u_k-u_k'\right) \right\rangle, \\\\
\ds G =  
\frac{\partial}{\partial r_k} \left\langle  \vartheta \vartheta' \left( u_k-u_k'\right) \right\rangle,
\end{array}
\label{KG}
\eea
where the repeated index denotes the summation convention.
$K$ and $G$ are linked to the longitudinal triple velocity correlation 
function $k$, and to the triple correlation between $u_r$ and $\vartheta$ 
following  
\bea
\begin{array}{l@{\hspace{+0.0cm}}l}
\ds K(r) = u^3 \left( \frac{\partial }{\partial r} + \frac{4}{r} \right) 
k(r), 
\ \ \mbox{where} \ \ 
\ds k(r) = \frac{\langle u_r^2 u_r' \rangle}{u^3}, \\\\
\ds G(r) = 2 u \theta^2 \left( \frac{\partial }{\partial r} + \frac{2}{r} 
\right) m^*(r), 
\ \ \mbox{where} \ \ 
\ds m^*(r) = \frac{\langle u_r \vartheta \vartheta' \rangle}{\theta^2 u},
\end{array}
\eea
Without particular hypotheses about the statistics of $\bf u$ and $\vartheta$,  $K$ and $G$ are unknown quantities \cite{Karman38, Corrsin_1, Corrsin_2} which can not be expressed in terms of $f$ and $f_\theta$, thus at this stage of this analysis, both the correlations equations (\ref{vk-h}) 
result to be unclosed.
To obtain analytical forms of $K$ and $G$, observe that these latter, representing the energy flow between length scales, do not modify the total amount of kinetic and thermal energies \cite{Batchelor53, Corrsin_1}.
On the other hand, the proposed statistical independence (\ref{indep stat}) allows to express ${\partial P}/{\partial t}$ as sum of two terms
\bea
\begin{array}{l@{\hspace{+0.0cm}}l}
\ds \frac{\partial P}{\partial t} = 
P_\xi \frac{\partial F}{\partial t} + 
F \frac{\partial P_\xi}{\partial t} 
\end{array}
\label{f_t}
\eea 
the first one of which, related to ${\partial F}/{\partial t}$, provides the time variations of velocity and temperature fields, whereas the second one, linked to ${\partial P_\xi}/{\partial t}$, not producing changing of $u^2$ and $\theta^2$, identifies the energy cascade effect. Therefore, $K$ and $G$ arise from the second term of (\ref{f_t}), and can be written, using the Liouville theorem (\ref{Liouville}) in terms of material displacements $\bfxi$, taking into account flow homogeneity and fluid incompressibility.
Specifically, from Eq. (\ref{Liouville}), $K$ and $G$ directly arise from 
$- F \partial(P_\xi \dot{\bfxi})/\partial{\bfxi}$ and are calculated as follows
\bea
\begin{array}{l@{\hspace{+0.0cm}}l}
\ds \left( 3+ r \frac{\partial}{\partial r} \right) K = - \int_{\cal U} \int_\Xi F \frac{\partial}{\partial {\bfxi}} \cdot \left( P_\xi \dot{\bfxi} \right) 
u_i u_i^* \ d{\cal U} d\Xi, \\\\
\ds G = - \int_{\cal U} \int_\Xi F \frac{\partial}{\partial {\bfxi}} \cdot \left( P_\xi \dot{\bfxi} \right) 
\ \vartheta \vartheta^* d{\cal U} d\Xi,
\end{array}
\label{KG1}
\eea
where $\ds {\cal U} = \left\lbrace {\bf u}\right\rbrace \times \left\lbrace {\vartheta}\right\rbrace$,  $\ds \Xi =  \left\lbrace {\bfxi}\right\rbrace$ and $d{\cal U}$ and $d\Xi$ are the corresponding elemental volumes, and
\bea
\begin{array}{l@{\hspace{-0.cm}}l}
\ds u_i = u_i(t, {\bfx}), \ \ u_i^* = u_i(t, {\bfx}+ {\bfxi}), \ \ i=1, 2, 3, \\\\
\ds \vartheta = \vartheta(t, {\bfx}), \ \ \vartheta^* = \vartheta(t, {\bfx}+ {\bfxi}), 
\end{array}
\eea
Integrating Eqs. (\ref{KG1}) with respect to ${\cal U}$, and considering that in incompressible isotropic turbulence $\left\langle u_i u_i' \right\rangle$ is related to $f$ through \cite{Batchelor53, Karman38}
\bea
\begin{array}{l@{\hspace{+0.0cm}}l}
\ds \left\langle u_i u_i^*  \right\rangle = u^2 \left( 3+ \xi \frac{\partial}{\partial \xi} \right) f(\xi) 
\end{array}
\eea
 we obtain 
\bea
\begin{array}{l@{\hspace{+0.0cm}}l}
\ds \left( 3+ r \frac{\partial}{\partial r} \right) K 
= - u^2 \int_\Xi \frac{\partial}{\partial {\bfxi}} \cdot \left( P_\xi \dot{\bfxi} \right) 
\left\langle u_i u_i^* \right\rangle \ d\Xi \\\\
\ds = - u^2 \int_\Xi \frac{\partial}{\partial {\bfxi}} \cdot \left( P_\xi \dot{\bfxi} \right) 
\left( 3+ \xi \frac{\partial}{\partial \xi} \right) f(\xi) \ d\Xi, \\\\
\ds G = - \theta^2 \int_\Xi \frac{\partial}{\partial {\bfxi}} \cdot \left( P_\xi \dot{\bfxi} \right) 
 f_\theta(\xi) \ d\Xi,
\end{array}
\label{KG2}
\eea
As $P_\xi$ is represented by a Dirac's delta distribution (\ref{Pxi_1}), thanks to 
Eq. (\ref{property}) (see the Appendix), to the fluid incompressibility and  to the isotropy, $K$ and $G$ read as  
\bea
\begin{array}{l@{\hspace{+0.0cm}}l}
\ds \left( 3+ r \frac{\partial}{\partial r} \right)  K =  u^2 \left( 3+ r \frac{\partial}{\partial r} \right)  \int_\Xi P_\xi \frac{\partial f}{\partial {\bfxi}} \cdot \dot{\bfxi}  
\ d\Xi \\\\
\ds = u^2 \left( 3+ r \frac{\partial}{\partial r} \right) \int_\Xi P_\xi \frac{\partial f}{\partial \xi} \frac{\bfxi}{\xi} \cdot \dot{\bfxi}  
\ d\Xi, 
\end{array}
\label{KG3}
\eea
\bea
\begin{array}{l@{\hspace{+0.0cm}}l}
\ds G = \theta^2 \int_\Xi P_\xi \frac{\partial f_\theta}{\partial {\bfxi}} \cdot \dot{\bfxi}  
 \ d\Xi = \theta^2 \int_\Xi P_\xi \frac{\partial f_\theta}{\partial \xi} \frac{\bfxi}{\xi} \cdot \dot{\bfxi}  
 \ d\Xi,
\end{array}
\label{KG3 bis}
\eea
Next, $K$ does not modify the average kinetic energy ($K(0)$=0), Eq. (\ref{KG3}) admits first integral which cancels $(3+r \partial/\partial r)$, thus $K$ and $G$ are 
\bea
\begin{array}{l@{\hspace{+0.0cm}}l}
\ds K =  u^2 \int_\Xi P_\xi \frac{\partial f}{\partial \xi} U  
\ d\Xi =u^2 \frac{\partial f}{\partial r}  \left\langle U \right\rangle_\xi, \\\\
\ds G = \theta^2 \int_\Xi P_\xi \frac{\partial f_\theta}{\partial \xi} U  
 \ d\Xi = \theta^2 \frac{\partial f_\theta}{\partial r}  
\left\langle U  \right\rangle_\xi,
\end{array}
\label{KG5}
\eea
Furthermore, the variance of $U$ is linked to the velocity correlation by means of the relationship
\bea
\begin{array}{l@{\hspace{+0.0cm}}l}
\ds \left\langle  U^2 \right\rangle_\xi \equiv  \left\langle (u_\xi^*-u_\xi)^2 \right\rangle_\xi = 2 u^2 \left( 1 - f(r) \right), \\\\
\ds \mbox{being} \ u_\xi = {\bf u}(t, {\bf x})\cdot \frac{\bfxi}{\xi}, \ \ u_\xi' = {\bf u}(t, {\bf x}+ 
{\bfxi})\cdot \frac{\bfxi}{\xi}  
\end{array}
\label{lambda f}
\eea
where $f$ is now calculated as the average of $u_\xi u_\xi^*/u^2$ through $P_\xi$ using Eq. 
(\ref{corr u 2}), and $\langle U \rangle_\xi$ and $\langle U^2 \rangle_\xi$ are related with each other through Eq. (\ref{mPl}).

This leads to the closure formulas of $K$ and $G$ in terms of autocorrelations and of their gradients
\bea
\begin{array}{l@{\hspace{+0.0cm}}l}
\ds K(r) = u^3 \sqrt{\frac{1-f}{2}} \frac{\partial f}{\partial r}, \\\\
\ds G(r) = u \theta^2 \sqrt{\frac{1-f}{2}} \frac{\partial f_\theta}{\partial r},
\end{array}
\label{K}
\label{K closure}
\eea
These closures, being not based on the eddy viscosity concept, do not exhibit second order derivatives of correlations, thus Eqs. (\ref{K}) do not represent a diffusive model.
These equations are the result of the trajectories divergence in the continuum fluid,
for which the mechanism of turbulence cascade consists in a propagation phenomenon of the pair correlations $f$ and $f_\theta$ between the several scales $r$ with a variable propagation speed $c_T$ which depends on $r$ and $u$, according to
\bea
\ds c_T = u \sqrt{\frac{1-f}{2}}
\eea 
The main asset of Eqs. (\ref{K closure}) with respect to the other closures is that such equations are not the result of  phenomenological assumptions, being these achieved by means of Liouville theorem and statistical independence of $\bfxi$ and $\bf u$ expressed by Eq. (\ref{indep stat}). 
It is worth to remark the importance of this latter.
Such equation, being the hypothesis of fully developed turbulence of this analysis, allows to
analytically express $K$ and $G$ separating the effects of the trajectories divergence in the physical space from those of the velocity field fluctuations in the Navier--Stokes phase space. 
Due to their theoretical foundation, Eqs. (\ref{K closure}) do not exhibit free model parameters or empirical constants which have to be identified.

These closures coincide with those just obtained by the author in the previous works \cite{deDivitiis_1, deDivitiis_4, deDivitiis_5, deDivitiis_8}.
While these latter derive such formulas exploiting, among the other things, the finite--scale Lyapunov theory, here, unlike the previous articles, Eqs. (\ref{K closure}) are obtained only using the Liouville theorem and the hypothesis of fully developed chaos, showing that the proposed closures do not need
the finite--scale Lyapunov theory, resulting this latter to be a sufficient theoretical tool for achieving the same formulas.

As regards the results obtained with the proposed closures, the reader is referred to the data presented in the previous works \cite{deDivitiis_1, deDivitiis_2, deDivitiis_4, deDivitiis_5, deDivitiis_8} 
In brief, we recall that Refs. \cite{deDivitiis_1,  deDivitiis_4, deDivitiis_5} and \cite{deDivitiis_2} show that such closures adequately describe the energy cascade phenomenon, reproducing,  negative skewness of velocity difference 
\bea
\ds H_3(r) \equiv 
\frac{\langle (\Delta u_r)^3 \rangle }{\langle (\Delta u_r)^2 \rangle^{3/2}} 
=
\frac{6 k(r)}{(2(1-f(r)))^{3/2}}
\label{H3}
\eea
such that
\bea
\ds H_3(0) = - \frac{3}{7},
\eea
in very good agreement with the litarature \cite{Chen92, Orszag72, Panda89, Anderson99, Carati95, Kang2003},  the Kolmogorov law and temperature spectra in line with the theoretical argumentation of Kolmogorov, Obukhov--Corrsin and Batchelor \cite{Batchelor_2, Batchelor_3, Obukhov}, with experimental results \cite{Gibson, Mydlarski}, and with numerical data \cite{Rogallo, Donzis}.
Furthermore, Ref \cite{deDivitiis_8} shows that the proposed closure formulas give a Kolmogorov constant of about 2, and produce correlations self-–similarity in proper interval of $r$, directly caused by the continuous fluid particles trajectories divergence.

We conclude this section  by observing the limits of the proposed closures (\ref{K}).
These limits directly derive from the hypotheses under which Eqs. (\ref{K}) are obtained: 
Eqs. (\ref{K}) are valid only in regime of fully developed chaos where the turbulence
exhibit homogeneity and isotropy. Otherwise, during the transition through intermediate stages of turbulence, or in more complex situations with particular boundary conditions, for instance in the presence of wall, Eqs. (\ref{K}) cannot be applied.

\bigskip

\section{Conditions for the correlation equations invariants and self--similarity}

{{ \color{black}

Using the proposed closures, the existence conditions of the correlation equations
invariant are here studied, with particular reference to the integrals of Loitsianskii and of Saffman--Birkhoff \cite{Loitsianskii39, Saffman67}.

Thereafter, given these invariants, the sole knowledge of the time evolution of ($u$, $\lambda_T$) and of ($\theta$, $\lambda_\theta)$ is shown to be a sufficient element to adequately describe the decay of isotropic homogeneous turbulence if the similarity of the spectra is self--preserved.

To study the first question, observe that the correlations equations admit, under certain conditions,
the following invariant integrals
\bea
\begin{array}{l@{\hspace{+0.0cm}}l}
\ds I_u = u^2 \int_0^\infty f r^4 dr, \\\\
\ds I_\theta =\theta^2 \int_0^\infty f_\theta r^2 dr, \\\\
\ds I_{SB} = \frac{u^2}{2} \int_0^\infty \left( 3 f + r \frac{\partial f}{\partial r} \right) r^2 dr,
\end{array}
\label{Saffman} 
\eea
where $I_u$ and $I_{SB}$ are, respectively, Loitsianskii integral  and Saffman--Birkhoff invariant \cite{Loitsianskii39, Saffman67}, two quantities associated with the von K\'arm\'an--Howarth equation based on angular momentum \cite{Landau59} and linear momentum, whereas $I_\theta$ is the invariant associated with the Corrsin equation.

Now, to determine the existence conditions of such invariants compatible with the proposed closures,
observe that the time evolutions of  $I_u$, $I_{SB}$ and $I_\theta$ are linked to $f$ and $f_\theta$  through the correlation equations, i.e. 
\bea
\begin{array}{l@{\hspace{+0.0cm}}l}
\ds \frac{d I_u}{dt} = u^3 \left[ r^4 k \right]_0^\infty +2 \nu u^2 \left[ r^4 \frac{\partial f}{\partial r}\right]_0^\infty, \\\\
\ds \frac{d I_\theta}{dt} = 2 u \theta^2 \left[ r^2 m^* \right]_0^\infty +2 \chi \theta^2 \left[ r^2 \frac{\partial f_\theta}{\partial r}\right]_0^\infty, \\\\
\ds \frac{d I_{SB}}{dt} = \frac{1}{2}\left[ r^3 K \right]_0^\infty + \nu u^2 \left[ r^2 \frac{\partial}{\partial r}
\left( 3 f + r \frac{\partial f}{\partial r} \right) \right]_0^\infty
\end{array}
\label{eq invar}
\eea
Without correlation equations closures, the existence condition of such invariants should be analyzed supposing that the asymptotic behaviors ($r \rightarrow \infty$) of $k$ and $f$  --as well as those of $m^*$ and $f_\theta$-- are independent. 
Here, assuming the proposed closures, double and triple correlations are related with each other according to
\bea
\begin{array}{l@{\hspace{+0.0cm}}l}
\ds K = u^3 \sqrt{\frac{1-f}{2}} \frac{\partial f}{\partial r} \equiv 
\frac{u^3}{r^4}\frac{\partial}{\partial r}\left( r^4 k\right), \\\\ 
\ds G = u \theta^2 \sqrt{\frac{1-f}{2}} \frac{\partial f_\theta}{\partial r} \equiv 
2 \frac{u \theta^2}{r^2}\frac{\partial}{\partial r}\left( r^2 m^*\right),  
\end{array}
\label{Kk}
\eea
As the result, the conditions for the existence of $I_u$ and $I_\theta$ must
prescribe that velocity correlations ($f$, $k$) and temperature correlations ($f_\theta$, $m^*$) tend to zero more rapidly than $r^{-4}$ and $r^{-2}$, respectively.
More in particular
\bea
\ds f \approx r^{-m}, 
\  r \rightarrow \infty,
\begin{array}{l@{\hspace{+0.0cm}}l}
    m > 4 \ \ \mbox{in general}, \\\\
  m \ge 5 \ \ \mbox{for} \ \vert I_u \vert < \infty
\end{array}
\label{invar1}
\ \mbox{for} \ \frac{d I_u}{dt}= 0,
\eea
\bea
\ds f_\theta \approx r^{-n},
 \ \ r \rightarrow \infty,
\begin{array}{l@{\hspace{+0.0cm}}l}
\ n > 2  \ \ \mbox{in general}, \\\\
\ n \ge 3 \ \ \mbox{for} \ \vert  I_\theta \vert < \infty,
\end{array}
\ \mbox{for} \ \frac{d I_\theta}{dt}= 0,
\label{invar2}
\eea
In fact, if $f \approx r^{-m}$ and $f_\theta \approx r^{-n}$ as $r \rightarrow \infty$, then
from Eq. (\ref{Kk}) $k\approx r^{-m} +c_u r^{-4}$ and $m^*\approx r^{-n} +c_\theta r^{-2}$,
where $c_u$ and $c_\theta$ are arbitrary constants. Therefore, assuming $c_u = c_\theta$=0, $m>$4 and $n>$2, both the terms at the R.H.S. of Eqs. (\ref{eq invar}) vanish and $I_u$ and $I_\theta$
maintain both their initial values during the decay. 
Viceversa, if $m\le$4, $n \le$2, the quantities $k r^4$ and $m^* r^2$ diverge as $r \rightarrow \infty$, and are different from zero if $m>$4 and $n>$2 with $c_u \ne$ 0, $c_\theta\ne$0, thus $I_u$ and $I_\theta$ will vary and can diverge during the  decay.
This agrees with Proudman and Reid \cite{Proudman54, Batchelor56} following which $I_u$ can be not an invariant as $\lim_{r \rightarrow \infty} k r^4$ is not in general equal to zero, and is also in line with Saffman \cite{Saffman67} which showed that $I_u$ can diverge in certain conditions.

As far as the Saffman--Birkhoff invariant is concerned, $I_{SB}$ is preserved 
if $f$ goes to zero more rapidly than $r^{-m}$, that is
\bea
\ds f \approx r^{-m}, \ \ r \rightarrow \infty,
\begin{array}{l@{\hspace{+0.0cm}}l}
m>2 \ \ \mbox{in general}, \\\\
m \ge 3 \ \ \mbox{for} \ \ \vert I_{SB} \vert < \infty, 
\end{array}
\ds \mbox{for} \ \frac{d I_{SB}}{dt}=0.
\label{invar3}
\eea
Equations (\ref{invar1}), (\ref{invar2}) and (\ref{invar3}) represent the conditions for which  $I_u$, $I_\theta$ and $I_{SB}$ are invariants, respectively, compatible with the proposed closures. 

Now, the existence of such invariants and the self--similarity produced
by the proposed closures, allow to show that the evolution of $u$, $\theta$, $\lambda_T$ and $\lambda_\theta$ can be sufficient to describe the isotropic turbulence decay.
To study this, the evolution equations of $u$, $\theta$, $\lambda_T$ and $\lambda_\theta$ are first deducted by taking the coefficients of order $r^0$ and $r^2$ of Eqs. (\ref{vk-h}) which arise from the Taylor series expansion of $f$ and $f_\theta$ \cite{Karman38}, \cite{Corrsin_1, Corrsin_2} 
\bea
\begin{array}{l@{\hspace{+0.0cm}}l}
\ds f= 1-\frac{1}{2}\left( \frac{r}{\lambda_T}\right)^2 +  \frac{1}{4 !} f^{IV}(t, 0) r^4+..., \\\\
\ds f_\theta= 1-\left( \frac{r}{\lambda_\theta}\right)^2 + \frac{1}{4 !} f_\theta^{IV}(t, 0) r^4+...,
\end{array}
\label{f f_theta}
\eea
This leads to the following equations
\bea
\begin{array}{l@{\hspace{+0.2cm}}l}
\ds \frac{d u^2}{d t} = - \frac{10 \nu}{\lambda_T^2} u^2, \\\\
\ds \frac{d \theta^2}{d t} = - \frac{12 \chi}{\lambda_\theta^2} \theta^2, 
\end{array}
\label{u2dot_thetadot}
\eea
\bea
\begin{array}{l@{\hspace{+0.0cm}}l}
\ds  \frac{d \lambda_T}{dt} = -\frac{u}{2} + \frac{\nu}{\lambda_T}
\left( \frac{7}{3} f^{IV}(t, 0) \lambda_T^4-5\right), \\\\
\ds  \frac{d \lambda_\theta}{dt} = -\frac{u}{2}\frac{\lambda_\theta}{\lambda_T} + \frac{\chi}{\lambda_\theta}
\left( \frac{5}{6} f^{IV}_\theta(t, 0) \lambda_\theta^4-6\right)
\end{array}
\label{lambda_dot} 
\eea
Equations (\ref{lambda_dot}) are not closed because incorporate $f^{IV}(t, 0)$ and $f^{IV}_\theta(t, 0)$ which in turn need an adequate estimation.
To fairly estimate these latter in function of $u$, $\theta$, $\lambda_T$ and $\lambda_\theta$, this analysis exploits the possible existence of the invariants $I_u$, $I_\theta$ and $I_{SB}$, and the properties that the proposed closures generate self--similarity of $f$ and $f_\theta$ in proper ranges of $r$ \cite{deDivitiis_8}. In such these intervals, velocity and temperature correlations can be approximated by
\bea
\begin{array}{l@{\hspace{+0.0cm}}l}
\ds f(t, r) \simeq f\left( \frac{r}{\lambda_T(t)}\right) , \\\\
\ds f_\theta(t, r) \simeq f_\theta\left( \frac{r}{\lambda_\theta(t)}\right) , 
\end{array}
\label{similar}
\eea
Hence, Eq. (\ref{similar}) provides the following relations, each for a single invariant
\bea
\begin{array}{l@{\hspace{+0.0cm}}l}
\ds u^2 \lambda_T^5 \simeq \mbox{const}, \ \  \mbox{if} \ I_u = \mbox{const}, \\\\
\ds u^2 \lambda_T^3 \simeq \mbox{const}, \ \  \mbox{if} \ I_{SB} = \mbox{const}, \\\\
\ds \theta^2 \lambda_\theta^3 \simeq \mbox{const} \ \ \mbox{if} \ I_\theta = \mbox{const}
\end{array}
\label{invars1}
\eea
Now, combining Eq. (\ref{invars1}) with Eqs. (\ref{u2dot_thetadot}) and (\ref{lambda_dot}),
one obtains a link between correlation scales and the fourth order derivatives of the correlations 
which holds for the proposed closure formulas
\bea
\begin{array}{l@{\hspace{+0.0cm}}l}
\ds f^{IV}(t,0) \lambda_T^4(t) = \frac{3}{14}\left( 14 +R_T \right),  \ \  \mbox{if} \ I_u = \mbox{const}, \\\\
\ds f^{IV}(t,0) \lambda_T^4(t) = \frac{3}{14}\left( \frac{50}{3} +R_T \right),  \ \  \mbox{if} \ I_{SB} = \mbox{const}, \\\\
\ds f^{IV}_\theta(t,0) \lambda_\theta^4(t) = \frac{3}{5}\left( 20 +R_T Pr \left( \frac{\lambda_\theta}{\lambda_T}\right)^2  \right), \ \  \mbox{if} \ I_\theta = \mbox{const},
\end{array}
\label{f4lambda}
\eea
being $R_T=u \lambda_T/\nu$ and $Pr = \nu/\chi$ Taylor scale Reynolds number, and Prandtl
number, respectvely. 
Hence, the evolution laws of $u$, $\lambda_T$, $\theta$, $\lambda_\theta$ are
\bea
\begin{array}{l@{\hspace{+0.0cm}}l}
\ds  \frac{\lambda_T^2(t)}{\lambda_T^2(0)}=1+ \frac{4 \nu t}{\lambda_T^2(0)}, \ \ 
\frac{u^2(t)}{u^2(0)}=\left( 1+ \frac{4 \nu t}{\lambda_T^2(0)} \right)^{-5/2},   \mbox{if} \ I_u = \mbox{const}, \\\\
\ds  \frac{\lambda_T^2(t)}{\lambda_T^2(0)}=1+ \frac{20}{3}\frac{\nu t}{\lambda_T^2(0)}, \ \
\frac{u^2(t)}{u^2(0)}=\left( 1+ \frac{20}{3}\frac{\nu t}{\lambda_T^2(0)} \right)^{-3/2},    \mbox{if} \ I_{SB} = \mbox{const}, \\\\
\ds \frac{\lambda_\theta^2(t)}{\lambda_\theta^2(0)}=1+ \frac{8 \chi t}{\lambda_\theta^2(0)}, \ \  \frac{\theta^2(t)}{\theta^2(0)}=\left( 1+ \frac{8 \chi t}{\lambda_\theta^2(0)} \right)^{-3/2},    \mbox{if} \ I_\theta = \mbox{const},
\end{array}
\eea
For what concerns $R_T$, it varies according to 
\bea
\begin{array}{l@{\hspace{+0.0cm}}l}
\ds \frac{R_T(t)}{R_T(0)}=\left( 1+ \frac{4 \nu t}{\lambda_T^2(0)} \right)^{-3/4},  \mbox{if} \ I_u = \mbox{const}, \\\\
\ds \frac{R_T(t)}{R_T(0)}=\left( 1+ \frac{20}{3}\frac{\nu t}{\lambda_T^2(0)} \right)^{-1/4},  \mbox{if} \ I_{SB} = \mbox{const}, 
\end{array}
\eea
For the proposed closures, in the presence of self--similarity and invariants, the sole evolution equations of ($u$, $\lambda_T$) and  ($\theta$, $\lambda_\theta$), which represent an ordinary differential system in homogeneous isotropic turbulence, can be considered to be sufficient to adequately describe the turbulence decay.
}}

\bigskip

\section{Appendix: Some properties of Dirac's delta distribution \label{properties}}

This section renews some of the more significant properties of the Dirac's delta distribution
\cite{Dirac58, Bracewell86, Arfken2000} useful to the purposes of the present analysis.

As well known, the Dirac's delta distribution, $\delta(x)$, is a generalized function obtained 
as limit of a class of strongly peaked functions \cite{Bracewell86}. 
The Dirac's delta exhibits the basic property 
\bea
\begin{array}{l@{\hspace{-0.cm}}l}
\ds  \int_{-\infty}^{\infty} \delta(x-x_0) f(x) \ dx  = f(x_0)
\label{definition}
\end{array}
\eea 
where $f$ is an arbitrary smooth function of $x$.
Among the other  properties, we mention the equations which involves the Dirac's delta,
its derivatives $\delta^{(h)}(x)$ and arbitrary differentiable functions $f$
\bea
\begin{array}{l@{\hspace{-0.cm}}l}
\ds  \int_{-\infty}^{\infty} \delta^{(h)}(x) f(x) \ dx  = 
- \int_{-\infty}^{\infty} \delta^{(h-1)}(x) \frac{\partial f}{\partial x} \ dx, \ \ h = 0, 1, 2, ...
\label{property 0}
\end{array}
\eea 
which lead to express the Dirac's delta derivatives in terms of $\delta$
\bea
\begin{array}{l@{\hspace{-0.cm}}l}
\ds   \delta^{(h)}(x)  =  \delta(x) \frac{(-1)^h h! }{x^h}, \ \ h= 0, 1, 2,...
\label{derivative}
\end{array}
\eea

In a n--dimensional space $S^n$, ${\bf x}\equiv(x_1, x_2, ... , x_n)$, the Dirac's delta is defined as
\bea
\ds \delta({\bf x}) = \ds \prod_{k=1}^n \delta(x_k)
\eea
If ${\bf f}\equiv(f_1, f_2, ... , f_n)$,  such this distribution and $\bf f$ satisfy the following identity
\bea
\begin{array}{l@{\hspace{-0.cm}}l}
\ds  \int_{-\infty}^{\infty} \int_{-\infty}^{\infty} ... \int_{-\infty}^{\infty} \nabla_{\bf x} \delta({\bf x}) \cdot {\bf f} \ dX  = 
-  \int_{-\infty}^{\infty} \int_{-\infty}^{\infty} ... \int_{-\infty}^{\infty} \delta({\bf x}) \nabla_{\bf x} \cdot {\bf f} \ dX, \\\\
\mbox{where} \ \ dX = dx_1 dx_2 ... dx_n, 
\end{array} 
\label{property}
\eea
$f_1$, $f_2$,...,$f_n$ are arbitrary smooth scalar functions of $\bf x$, and $\cdot$ and $\nabla_{\bf x} \cdot$, stand for respectively, inner product and divergence operator,
both defined in $S^n$.

\bigskip 

\section{Conclusion \label{Conclusion}}

The closure formulas of von K\'arm\'an--Howarth and Corrsin equations are achieved
using Liouville theorem and rate of the material lines distribution function in isotropic homogeneous
turbulence. These closures coincide with those given in previous 
works which adopt also the finite--scale Lyapunov analysis.
The present work corroborates the previous results, giving a proof of the closures more general and more rigorous than that presented in the previous works, and showing that such closures do not need the Lyapunov analysis, the latter being a sufficient theoretical element that leads to the same formulas.
{\color{black} 
Furthermore, this study analyzes the existence conditions of invariants of the closed correlation equations, and shows that, in the presence of self--similarity and such invariants, the isotropic turbulence can be fairly described through the sole evolution of ($u$, $\lambda_T$) and  ($\theta$, $\lambda_\theta$), compatible with the present closure formulas.
}

\bigskip 

\section{Acknowledgments}

This work was partially supported by the Italian Ministry for the Universities 
and Scientific and Technological Research (MIUR). 

\bigskip

\end{document}